\documentclass[11pt]{article}
\usepackage{fullpage}
\usepackage{times}

\usepackage{doi}

\usepackage{graphicx}
\usepackage{listings}

\usepackage{amsmath}
\usepackage{amsfonts}
\usepackage{amssymb}
\usepackage{amsthm}
\theoremstyle{definition}

\usepackage{xspace}

\usepackage{prettyref}
\newcommand{\cref}[2][]{\prettyref{#2}}
\newrefformat{sec}{Sec.\,\ref{#1}}
\newrefformat{subsec}{Sec.\,\ref{#1}}
\newrefformat{def}{Def.\,\ref{#1}}
\newrefformat{thm}{Theorem\,\ref{#1}}
\newrefformat{prop}{Proposition\,\ref{#1}}
\newrefformat{rem}{Remark\,\ref{#1}}
\newrefformat{lem}{Lemma\,\ref{#1}}
\newrefformat{cor}{Corollary\,\ref{#1}}
\newrefformat{ex}{Example\,\ref{#1}}
\newrefformat{cex}{Counterexample\,\ref{#1}}
\newrefformat{tab}{Table\,\ref{#1}}
\newrefformat{fig}{Fig.\,\ref{#1}}
\newrefformat{case}{case\,\ref{#1}}
\newrefformat{foot}{Footnote\,\ref{#1}}

\usepackage{mathtools}
\usepackage{tikz}
\usepackage{amssymb}
\usepackage{mathbbol} % lowercase blackboard bold
\usepackage{stmaryrd}
\usepackage[T1]{fontenc}
\usepackage{multicol}

\usepackage{tikz,pgfplots,pgfplotstable}
\usetikzlibrary{patterns}
\pgfplotsset{compat=1.18}

\usepackage{color} % colors referenced in lstlean.tex
\definecolor{keywordcolor}{rgb}{0.7, 0.1, 0.1}   % red
\definecolor{commentcolor}{rgb}{0.4, 0.4, 0.4}   % grey
\definecolor{symbolcolor}{rgb}{0.0, 0.1, 0.6}    % blue
\definecolor{sortcolor}{rgb}{0.1, 0.5, 0.1}      % green

\usepackage{mdframed}

\usepackage{booktabs}

\usepackage{xspace}

\usepackage{MnSymbol}

\usepackage{listings}
\usepackage{xcolor}
\usepackage{siunitx}

\newcommand{\ttt}{\texttt}
\newcommand{\tit}{\textit}
\newcommand{\tbf}{\textbf}

\newcommand{\isa}{\lstinline[language=isabelle]}
\newcommand{\iisa}{\lstinline[language=isabelle2]}

\newcommand{\toolname}{\isa{Apply2Isar}\xspace}

\definecolor{ctblue}{HTML}{006699}
\definecolor{cbblue}{HTML}{009966}
\definecolor{isarred}{HTML}{AF1010}
\definecolor{isarred2}{HTML}{DF1010}
\definecolor{isargreen}{HTML}{009966}
\definecolor{isarblue}{HTML}{3D54D1}
\definecolor{isarblue2}{HTML}{3E55D6}
\definecolor{isarblue3}{HTML}{0FA7FD}
\definecolor{isarseagreen}{HTML}{00AA99}
\definecolor{isarviolett}{HTML}{A020F0}
\definecolor{smtviolett}{HTML}{660066}
\definecolor{smtblue}{HTML}{000088}
\definecolor{smtgreen}{HTML}{008800}
\definecolor{isarbase}{HTML}{000000}
\definecolor{isarcomment}{HTML}{13952D}
\definecolor{isaroption}{HTML}{730081}
\definecolor{isarstring}{HTML}{202500}
\definecolor{isarbackground}{HTML}{EEEEEE}

\lstdefinelanguage{isabelle}{%
	backgroundcolor = \color{isarbackground},
    comment = [l]{(*},
	commentstyle = \color{isarcomment},
	basicstyle=\small\color{isarbase}\ttfamily,
    alsoletter= . ? ' -,
	keywords=[1]{type_synonym,consts,datatype,fun,abbreviation,definition,lemma,theorem,corollary,thm,schematic_goal,
		by,using,unfolding,have,moreover,ultimately,also,finally,from,note,proof,qed,.,..,oops},
	keywordstyle=[1]\bfseries\color{isarblue},
	keywords=[2]{and,if,where,assumes,shows,fixes,defines,for,premises,in,induct},
	keywordstyle=[2]\bfseries\color{isargreen},
	keywords=[3]{ite,then,else,case,AND,mod,unat,len,LENGTH,bit,False,sint,pow2},
	keywordstyle=[3]\color{isarblue2},
    keywords=[4]{show,assume,fix},
	keywordstyle=[4]\bfseries\color{isarblue3},
	keywords=[5]{parse_rare_file},
	keywordstyle=[5]\bfseries\color{isarblue},
    keywords=[6]{apply,prefer,defer,subgoal,done,back,supply,apply_end},
	keywordstyle=[6]\bfseries\color{isarred},
	keywords=[7]{sorry},
	keywordstyle=[7]\bfseries\color{isarred2},
    keywords=[8]{-,THEN,of,OF,metis,rule,fact,simp,auto,erule,assumption,standard,simp_all,safe,meson,unfold_locales,
	m,
	m1,
	m2,
	m3,
	m4,
	m5,
	m6,
	m7,
	m8,
	m9,
	m10,
	m25,
	m26,
	???
	},
    keywordstyle=[8]\bfseries\color{isarseagreen},
	keywords=[9]{add},
	keywordstyle=[9]\bfseries\color{isarviolett},
    keywords=[10]{tracing,print_types,smart_goals,smart_unfolds,named_facts,split_subgoals,
	split_fact_tac,dummy_subproofs,subgoal_fix_fresh,fact_name_prefix},
    keywordstyle=[10]\bfseries\color{isaroption},
	emph={'a, nat, string, word, bool},
    emphstyle=\color{isarviolett},
	escapeinside={<@}{@>},
	columns=fixed,
	extendedchars,
	basewidth={0.5em,0.45em},
	stringstyle=\ttfamily\color{isarviolett},
	upquote=true,
	showstringspaces=true,
	literate=
	{∀}{$\forall$}1
	{∃}{$\exists$}1
	{->}{{$\rightarrow$}}1
	{<-}{{$\leftarrow$}}1
	{<=}{{$\le$}}1
	{≤}{{$\le$}}1
	{>=}{{$\ge$}}1
	{≥}{{$\ge$}}1
	{<->}{{$\leftrightarrow$}}1
	{-->}{{$\longrightarrow$}}3
	{⟶}{{$\longrightarrow$}}3
	{<-->}{{$\longleftrightarrow$}}1
	{=>}{{$\Rightarrow$}}1
	{==}{{$\equiv$}}2
    {≡}{{$\equiv$}}2
	{==>}{{$\Longrightarrow$}}4
	{⟹}{{$\Longrightarrow$}}4
	{<=>}{{$\Leftrightarrow$}}1
    {‹}{{\guilsinglleft}}1
    {›}{{\guilsinglright}}1
    {∧}{{$\land$}}1
    {∨}{{$\lor$}}1
    {⋀}{{$\bigwedge$}}2
    {⟦}{{$\llbracket$}}2
    {[|}{{$\llbracket$}}2
    {⟧}{{$\rrbracket$}}2
    {|]}{{$\rrbracket$}}2
	{andFixes}{{ {\rm\color{isargreen}\bfseries and} }}3
	{parseCommand}{{{\rm\bfseries\color{isarblue}parse\_rare\_file}}}{11}
}

\lstdefinelanguage{isabelle2}{%
	backgroundcolor = \color{isarbackground},
    comment = [l]{(*},
	commentstyle = \color{isarcomment},
	basicstyle=\small\color{isarbase}\ttfamily,
    alsoletter= ?,
	keywords=[1]{type_synonym,datatype,fun,abbreviation,definition,lemma,theorem,corollary,schematic_goal,
		by,using,unfolding,have,moreover,ultimately,also,finally,from,note,proof,qed,oops,
		consts},
	keywordstyle=[1]\bfseries\color{isarblue},
	keywords=[2]{if,and,where,assumes,shows,fixes,defines,for,premises},
	keywordstyle=[2]\bfseries\color{isargreen},
	keywords=[3]{ite,then,else,case,AND,mod,unat,len,LENGTH,bit,False,sint,pow2},
	keywordstyle=[3]\color{isarblue2},
    keywords=[4]{show,assume,fix},
	keywordstyle=[4]\bfseries\color{isarblue3},
	keywords=[5]{parse_rare_file},
	keywordstyle=[5]\bfseries\color{isarblue},
    keywords=[6]{apply,prefer,defer,subgoal,done,back,supply,apply_end},
	keywordstyle=[6]\bfseries\color{isarred},
	keywords=[7]{sorry},
	keywordstyle=[7]\bfseries\color{isarred2},
    keywords=[8]{THEN,of,metis,rule,fact,simp,auto,erule,assumption,unfold_locales
	tactic,
	tactic1,
	tactic2,
	tactic3,
	tactic4,
	tactic5,
	tactic6,
	tactic7,
	tactic8,
	tactic9,
	tactic10,
	tactic25,
	tactic26,
	???
	},
    keywordstyle=[8]\bfseries\color{isarseagreen},
	keywords=[9]{add},
	keywordstyle=[9]\bfseries\color{isarviolett},
    keywords=[10]{tracing,print_types,smart_goals,smart_unfolds,named_facts,split_subgoals,
	split_fact_tac,dummy_subproofs},
    keywordstyle=[10]\bfseries\color{isaroption},
	emph={'a, nat, string, word, bool},
    emphstyle=\color{isarviolett},
	escapeinside={<@}{@>},
	columns=fixed,
	extendedchars,
	basewidth={0.5em,0.45em},
	stringstyle=\ttfamily\color{isarviolett},
	upquote=true,
	showstringspaces=true,
	literate=
	{∀}{$\forall$}1
	{->}{{$\rightarrow$}}1
	{<-}{{$\leftarrow$}}1
	{<=}{{$\le$}}1
	{≤}{{$\le$}}1
	{>=}{{$\ge$}}1
	{≥}{{$\ge$}}1
	{<->}{{$\leftrightarrow$}}1
	{-->}{{$\longrightarrow$}}3
	{⟶}{{$\longrightarrow$}}3
	{<-->}{{$\longleftrightarrow$}}1
	{=>}{{$\Rightarrow$}}1
	{==}{{$\equiv$}}2
    {≡}{{$\equiv$}}2
	{==>}{{$\Longrightarrow$}}4
	{⟹}{{$\Longrightarrow$}}4
	{<=>}{{$\Leftrightarrow$}}1
    {‹}{{\guilsinglleft}}1
    {›}{{\guilsinglright}}1
    {∧}{{$\land$}}1
    {∨}{{$\lor$}}1
    {⋀}{{$\bigwedge$}}2
	{andFixes}{{ {\rm\color{isargreen}\bfseries and} }}3
	{parseCommand}{{{\rm\bfseries\color{isarblue}parse\_rare\_file}}}{11}
}

\lstdefinelanguage{isabelle3}{%
	backgroundcolor = \color{isarbackground},
    comment = [l]{(*},
	commentstyle = \color{isarcomment},
	basicstyle=\small\color{isarbase}\ttfamily,
    alsoletter= . ? ',
	keywords=[1]{type_synonym,consts,datatype,fun,abbreviation,definition,lemma,theorem,corollary,thm,
		using,unfolding,have,moreover,ultimately,also,finally,from,note,proof,qed,.,..,oops},
	keywordstyle=[1]\bfseries\color{isarblue},
	keywords=[2]{and,if,where,assumes,shows,fixes,defines,for,premises},
	keywordstyle=[2]\bfseries\color{isargreen},
	keywords=[3]{ite,then,else,case,AND,mod,unat,len,LENGTH,bit,False,sint,pow2},
	keywordstyle=[3]\color{isarblue2},
    keywords=[4]{show,assume,fix},
	keywordstyle=[4]\bfseries\color{isarblue3},
	keywords=[5]{parse_rare_file},
	keywordstyle=[5]\bfseries\color{isarblue},
    keywords=[6]{apply2isar},
	keywordstyle=[6]\bfseries\color{isarred},
	keywords=[7]{sorry},
	keywordstyle=[7]\bfseries\color{isarred2},
    keywords=[8]{THEN,of,OF,metis,rule,fact,simp,erule,assumption,standard},
    keywordstyle=[8]\bfseries\color{isarseagreen},
	keywords=[9]{add},
	keywordstyle=[9]\bfseries\color{isarviolett},
    keywords=[10]{tracing,print_types,smart_goals,smart_unfolds,named_facts,split_subgoals,
	split_fact_tac,dummy_subproofs,subgoal_fix_fresh},
    keywordstyle=[10]\bfseries\color{isaroption},
	emph={'a, nat, string, word, bool},
    emphstyle=\color{isarviolett},
	escapeinside={<@}{@>},
	columns=fixed,
	extendedchars,
	basewidth={0.5em,0.45em},
	stringstyle=\ttfamily\color{isarviolett},
	upquote=true,
	showstringspaces=true,
	literate=
	{∀}{$\forall$}1
	{∃}{$\exists$}1
	{->}{{$\rightarrow$}}1
	{<-}{{$\leftarrow$}}1
	{<=}{{$\le$}}1
	{≤}{{$\le$}}1
	{>=}{{$\ge$}}1
	{≥}{{$\ge$}}1
	{<->}{{$\leftrightarrow$}}1
	{-->}{{$\longrightarrow$}}3
	{⟶}{{$\longrightarrow$}}3
	{<-->}{{$\longleftrightarrow$}}1
	{=>}{{$\Rightarrow$}}1
	{==}{{$\equiv$}}2
    {≡}{{$\equiv$}}2
	{==>}{{$\Longrightarrow$}}4
	{⟹}{{$\Longrightarrow$}}4
	{<=>}{{$\Leftrightarrow$}}1
    {‹}{{\guilsinglleft}}1
    {›}{{\guilsinglright}}1
    {∧}{{$\land$}}1
    {∨}{{$\lor$}}1
    {⋀}{{$\bigwedge$}}2
    {⟦}{{$\llbracket$}}2
    {[|}{{$\llbracket$}}2
    {⟧}{{$\rrbracket$}}2
    {|]}{{$\rrbracket$}}2
	{andFixes}{{ {\rm\color{isargreen}\bfseries and} }}3
	{parseCommand}{{{\rm\bfseries\color{isarblue}parse\_rare\_file}}}{11}
}

\newcommand{\prove}{\emph{prove}~}
\newcommand{\state}{\emph{state}~}

\title{Apply2Isar: Automatically Converting Isabelle/HOL Apply-Style Proofs to Structured Isar}
\author{
Sage Binder
\thanks{University of Iowa, Iowa City, IA, USA,
\newline \{sage-binder, katherine-kosaian\}@uiowa.edu}
\and
Hanna Lachnitt
\thanks{Stanford University, Stanford, CA, USA,
\newline lachnitt@cs.stanford.edu}
\and
Katherine Kosaian
\footnotemark[1]
}

\usepackage{ifpdf}
\ifpdf
\fi

\date{}

\begin{document}
\maketitle
\begin{abstract}
In Isabelle/HOL,
declarative proofs written in the Isar language
are widely appreciated for their readability and robustness.
However,
some users may prefer writing procedural ``apply-style'' proofs
since they enable rapid exploration of the search space.
To get the best of both worlds,
we introduce Apply2Isar,
a tool for Isabelle/HOL that automatically converts apply-style proofs
to declarative Isar.
This allows users to write complex,
possibly fragile apply-style proofs,
and then automatically convert them
to more readable and robust declarative Isar proofs.
To demonstrate the efficacy of Apply2Isar in practice,
we evaluate it on a large benchmark set consisting of apply-style proofs
from the Isabelle Archive of Formal Proofs.
\end{abstract}

\section{Introduction}\label{sec:intro}
The interactive theorem prover Isabelle \cite{isabelle-next-700} allows two primary styles
for constructing proofs:
apply-style proof scripts
and structured Isar proofs.
Apply-style proofs are \tit{procedural} proofs that
generally work backward,
starting from the desired goal(s) and invoking methods
to either discharge open goals or produce new proof obligations---this essentially builds a proof tree.
Importantly, in an apply-style proof,
the user must interactively step through the proof to view
the intermediate proof states.
Structured Isar, on the other hand, is \tit{declarative}
in that the user explicitly writes intermediate goals
and the proof body generally moves forward
(though one can provide an initial backward reasoning step),
referencing previous goals to prove new goals.
Moreover,
structured Isar supports various proof structures
that allow complex non-linear chains of reasoning to be presented in a
more organized and human-readable way.
This improved expressivity is one reason that it is now generally preferred over apply-style proofs
by the Isabelle community,
including by the \tit{Archive of Formal Proofs} (AFP),\footnote{\url{https://isa-afp.org/}}
which is a large and actively-maintained online repository of Isabelle proof developments.

Isar's declarative format was designed with readability and flexibility in mind,
drawing inspiration from informal mathematical traditions and
high-level programming languages \cite{isabelle-isar},
as well as the declarative language of the Mizar proof assistant \cite{mizar-overview}.
It succeeds on many fronts:
explicit intermediate goals help clarify the proof outline and flow;
the forward reasoning is often clearer to follow;
and the multitude of reasoning patterns
allow proofs to more closely match the reasoning one might
find in a mathematical text.
Additionally,
compared to apply-style proofs,
structured Isar arguably provides inherent robustness benefits:
its declarative nature ensures that a proof's overall structure,
once written,
remains fixed even when method behaviors change
(which can happen, for instance, when simplification rules are added or removed,
or when a referenced lemma is refactored).
If a structured Isar proof does break,
it is usually easy to identify and repair broken steps,
even for a user who is unfamiliar with the details of the proof (such as an AFP editor).\footnote{
Isabelle/HOL's powerful automation tools
(like \isa{try0} and \isa{sledgehammer})
are often extremely effective at fixing broken steps in structured proofs.}

In contrast to structured Isar,
apply-style proofs can be difficult to repair:
if a step in an apply-style proof suddenly yields a different (unintended) proof state,
an error may not occur until much later in the proof,
which can make it hard to identify which method invocation is the culprit.
Even once the problematic step is identified,
how it should be repaired may not be clear
because the correct (intended) proof state is not a fixed property of the proof.

Despite the advantages of structured Isar,
apply-style proof are sometimes easier to \tit{write}
since they directly manipulate the proof state and let Isabelle 
automatically manage current goals,
simplify expressions,
apply lemmas with unification,
and so on.
Apply-style proofs also enable rapid exploration of the search space
because different methods can quickly be tried
with immediately visible results.
We explain the distinction between apply-style proofs and structured Isar in more depth
in \cref{sec:prelims}.

Even in structured Isar proofs,
Isabelle/HOL's Sledgehammer automation sometimes performs better
on goals that have first been simplified by \isa{simp}
or \isa{auto} \cite{hammering-away},
which can lead to brittle proof patterns if not refactored.
Thus, a user might want to write an apply-style proof first,
then convert it to structured Isar for greater readability.
To do a direct translation,
the user must step through the apply-style proof and
copy each set of intermediate goals
along with the corresponding method application.
Alternatively,
the user may use the insight gained from writing the apply-style proof
to write a new structured Isar proof from scratch.
Both methods are labor-intensive.
To alleviate this labor and achieve the best of both worlds,
we introduce \toolname,
a tool implemented in Isabelle/ML that
automatically converts Isabelle apply-style proofs to structured Isar proofs.
We give a motivating example of the potential brittleness of apply-style proofs
and an overview of \toolname in \cref{sec:motivating-example}.
As apply-style proofs can admit complex behavior that does not always translate neatly to structured Isar,
we encountered various design challenges in implementing \toolname,
which we discuss in \cref{sec:implementation}.
To demonstrate the efficacy of \toolname in practice,
we evaluated it on thousands of proofs from the AFP,
where it produced
total or partial translations in 95\%--99\% of test cases.
See \cref{sec:evaluation} for a detailed discussion of this evaluation.
In \cref{sec:future-and-related-work},
we discuss future work,
including how ideas from related proof refactoring tools may be integrated into \toolname.
We conclude in \cref{sec:conclusion}.

\section{Preliminaries}\label{sec:prelims}
The Isabelle tutorial provides detailed explanations
of the various proof commands \cite{isabelle-tutorial},
and the Isar reference manual explains the language in depth \cite{isabelle-isar-reference-manual}.
We review only basic concepts here,
and introduce further concepts throughout the paper as needed.

Isar is a proof language that enables users to write sequences of proof commands
in an interactive manner.
Methods can be used by commands to operate on pending goals.
New methods can be built up from other methods via combinators.
Methods do not allow direct goal addressing,
but either refer to
the first goal or to all goals.\footnote{
Although,
there is a combinator that applies a method to the first $n$ goals.}
While seemingly similar to methods,
tactics are programs written directly in ML,
the programming language in which Isabelle is implemented.
In the early days of Isabelle, the user would use ML functions,
including tactics, directly.
Nowadays,
tactic scripts are a thing of the past for
end-users who only interact with the Isar environment.

The Isar virtual machine (Isar/VM) keeps track of nodes consisting of the pending goal,
the proof context, and the linguistic mode.
There are two such modes that are of particular interest to us.
The first
is the \prove mode, which allows the user to refine goals,
e.g.,
with the \isa{apply m} command, which applies the proof method \isa{m}
and returns zero or more new goals.
The second mode
is the \state mode,
in which declarative statements 
such as new claims,
assumptions, or definitions can be written.
One way to do this is with the \isa{have [stmt]} command,
which opens a local goal and switches to the \prove mode.
Once all pending goals are proven,
the VM can switch from a node in the \prove mode to one in the \state mode,
e.g.,
via the \isa{done} command which finishes a proof.
For simplicity,
we omit any discussion of the third mode,
called the \emph{chain} mode.

For the purposes of this paper, we define an \tit{apply-style proof} to be one
where nodes are in the \prove mode,
which also classifies it as a procedural proof.
The only exception to this are steps that close a goal.
The name ``apply-style'' is somewhat of a misnomer since many other commands,
such as \isa{using} or \isa{unfolding},
can be used to transition between two nodes in the \prove mode.
However,
\isa{apply}
is often the most frequent command in these procedural proofs,
so Isabelle users often use the term ``apply-style''
to informally describe this style.
Apply-style proofs are an emulation of tactic scripts.
Note that users of other ITPs (and even many Isabelle users) will often refer
these kinds of procedural proofs as ``tactic scripts''.

The \isa{proof m} command can be used to transition from the \prove mode into the \state mode
after applying method \isa{m}.
The \isa{proof} command can also be invoked with a dash (\isa{proof(-)}) instead of a method,
in which case it directly transitions to the \state mode.
To finish a pending goal established by \isa{proof m},
\isa{show} is used in place of \isa{have};
if multiple goals are established by \isa{proof m},
each one must be proven in a \isa{show} statement.
Finally, the \isa{qed} command finishes the proof,
resulting in a new theorem.
This is the basis for what we will call a \tit{structured Isar proof} in this paper,
an instance of the declarative proof style.
While each new goal declaration will allow the user to switch back to apply-style proving,
we restrict structured Isar proofs to mean only those that
may use \isa{using} or \isa{unfolding} followed by a single
application of the \isa{by}, immediate proof (\isa{.}), or standard proof (\isa{..}) commands.
This enforces that for any complex transformation,
the goal must be explicitly stated.

In structured Isar proofs, facts can be labeled (e.g. \isa{have myFact: "X"})
and once proven, referenced later,
either by name (e.g. \isa{myFact}) or as a fact literal (e.g. \isa{‹X›}).
To avoid excessive fact labels,
one can chain facts, for instance with the \isa{then} and \iisa{moreover-ultimately} commands.
Additionally, one can fix local variables with \isa{fix}
and assert assumptions with \isa{assume};
the Isabelle kernel will refuse to prove a theorem if
any fixed variables,
asserted assumptions,
or \isa{show} statements do not match the goal(s).
The command \isa{by m} is similar to \isa{apply m done},
except it is \emph{terminal} in that it runs \isa{m} until it finds a solved state or fails.
One can also write \isa{by (m1) m2},
which tries to solve the current proof state
by applying \isa{m1} and \isa{m2} in sequence.
We explain more commands in Table \ref{tab:command-list}.

The following is an example of a structured Isar proof,
where we use \isa{[...]} to elide some of the commands used to discharge proof obligations:

\begin{lstlisting}[language=isabelle]
lemma "A ∧ C"
proof(-)
  have "A ∧ B" [...]                      (* Some proof of [ A ∧ B ] *)
  then have "A" by (rule conjunct1)       (* Uses the previous fact [ A ∧ B ] *)
  moreover have "C" [...]                 (* Some proof of [ C ] *)
  ultimately show "A ∧ C" by (rule conjI) (* Uses the collected facts [ A, C ] *)
qed
\end{lstlisting}

Throughout the paper,
we present many examples of apply-style and structured Isar proofs.
As the proof state is not explicit in apply-style proofs,
we use comments (denoted by \isa{(* ... *)})
to describe relevant information about the state (goals, facts in the context, etc.)
at key steps.
Note that when we write ``\isa{apply m (* ... *)}'',
we intend the comment to indicate information about the proof state
\tit{after} \isa{m} is applied.
We also use the common Isabelle convention of indenting each proof command
by $n - 1$ spaces beyond the baseline indentation of $2$,
where $n$ is the number of goals in the proof state when the command is invoked.

\section{Motivating Example and Overview}\label{sec:motivating-example}
Consider the following simple example of a lemma with an apply-style proof:
\begin{lstlisting}[language=isabelle]
consts A :: "bool" B1 :: "bool" B2 :: "bool" C :: "bool"

lemma L1: "B1 ⟹ B2 ⟹ A" [...]
lemma L2: "B1" [...]
lemma L3: "C ⟹ B2" [...]
lemma L4: "C" [...]

lemma apply_prf: "A" (* Our initial goal is [ A ] *)
  apply (rule L1)  (* By backward chaining, L1 reduces this goal to [ B1, B2 ] *)
   apply (rule L2) (* Applying L2 discharges B1, so [ B2 ] is the remaining goal *)
  apply (rule L3)  (* By backward chaining, L3 reduces this goal to [ C ] *)
  by (rule L4)     (* Applying L4 finally discharges [ C ] *)
\end{lstlisting}
Now,
suppose that lemma \isa{L1} is refactored
so its hypotheses are swapped:
\begin{lstlisting}[language=isabelle]
lemma L1: "B2 ⟹ B1 ⟹ A"  [...]
\end{lstlisting}
This simple change breaks the apply-style proof of lemma \isa{apply_prf} shown above.
Specifically,
\isa{rule L2} will now be applied to the goal \isa{B2} instead of \isa{B1},
resulting in an error.
A user investigating the broken step interactively can see that when \isa{rule L2} is applied,
the current goals are \isa{[ B2, B1 ]}.
However,
it is neither explicit that \isa{rule L2} failed because
it was expecting the first goal to be \isa{B1},
nor that the expected result of \isa{rule L2} is the goal \isa{[ B2 ]}.\footnote{
Isabelle includes options such as \isa{unify_trace} that,
when enabled, can help debug these apply-style proofs,
but we emphasize that this requires further manual interaction and some expertise---our point
is that the source of the error is not
immediately evident from the error message and proof text.
}

Now consider a corresponding structured Isar proof for the same lemma:
\begin{lstlisting}[language=isabelle]
lemma structured_prf: "A"
proof(-)
  have "B1" by (rule L2)
  moreover have "B2"
  proof(-)
    have "C" by (rule L4)
    then show "B2" by (rule L3)
  qed
  ultimately show "A" by (rule L1)
qed
\end{lstlisting}
If, in this structured proof, \isa{L1} is again changed as described above, 
the proof will fail at \isa{rule L1}---this more accurately reflects the cause of the failure.
Moreover,
it is now explicit that \isa{rule L1}
was intended to prove \isa{A} from the fact set \isa{[ B1, B2 ]}.
This helps the user repair the broken step or---as might be necessary
when the method application is very complicated---find an alternative proof.
Crucially,
the effects of the broken step are isolated---the still-working parts of the proof can remain intact.

\toolname
can automatically translate existing apply-style proofs to 
more maintainable structured Isar proofs.
At a high level,
\toolname replays a given apply-style proof
and records the remaining goals after each method application,
then constructs a structured Isar proof
by printing the intermediate goals (and the corresponding method applications)
in reverse.
The reversal is natural because Isar
is more conducive to forward reasoning whereas apply-style proofs generally proceed backward.
In the generated Isar,
we use the \isa{fact} method as ``glue'' between steps---the command \isa{by (fact h)} proves a goal that unifies with \isa{h},
and fails otherwise \cite{isabelle-isar-reference-manual}.

The previous translation example
demonstrates stylistic choices such as reordered goals
and the use of constructs
like subproofs and \iisa{moreover-ultimately}.
Used appropriately,
these choices can make the proof more natural to a human reader.
However, what makes a proof natural and readable is ultimately subjective.
Indeed,
readers with Isabelle/HOL expertise may have opinions on how the previous example
could have been more natural.
With \toolname,
we avoid making any stylistic choices on behalf of the user;
instead, we focus on producing a faithful translation
that corresponds closely to the original proof.
The faithful translation still benefits from the inherent readability and robustness
of structured Isar,
and it can also serve as a starting point for further readability improvements
made either by the user or by additional tools.
Notably,
our translation approach explicitly clarifies the logical flow of the original proof,
which we believe is useful information for additional refactoring passes.
We discuss potential extensions of our work in this direction in \cref{sec:future-and-related-work}.

Given the apply-style proof of lemma \isa{apply_prf} presented above,
\toolname produces the following structured proof:
\begin{lstlisting}[language=isabelle]
proof(-)
  have h_3_1: "C" by (rule L4)
  have h_2_2: "B2" by (rule L3) (fact h_3_1) 
  have h_2_1: "B1" by (rule L2)
  show h_1_1: "A" by (rule L1) (fact h_2_1, fact h_2_2) 
qed
\end{lstlisting}
One benefit of structured Isar is that it can express complex fact dependencies
in an organized and explicit way.
In the translated proof,
each goal has been labeled (in the form ``\isa{h_x_y}'')
so that it can be referenced later in the proof.\footnote{
Note that if these fact labels shadow existing lemma names,
the proof may break.
Thus,
we provide the \isa{fact_name_prefix} setting
to allow users to change the ``\isa{h}'' prefix to a custom string.
}
This is necessary because some methods can create new goals.
For instance,
\isa{rule L1} transforms the goal \isa{[ A ]} to the goals \isa{[ B1, B2 ]}.
So, to prove \isa{h_1_1: "A"} in the Isar proof,
we first apply \isa{rule L1},
then discharge the resulting goals \isa{B1} and \isa{B2} with \isa{fact h_2_1} and \isa{fact h_2_2}
respectively.

For users who find the fact labels too verbose,
we provide the \isa{named_facts} option.
If disabled,
no labels are generated.
Helpfully,
the \isa{fact} method can be invoked with no arguments,
in which case it automatically proves the current goal by
searching the context for a matching fact.
Thus,
when \isa{named_facts} is disabled,
we instead invoke \isa{fact+} to discharge the resulting goals after each method application
(the \isa{+} combinator applies a method one or more times).
To avoid clutter and unnecessary detail,
we present the rest of the examples in this paper without fact labels.

We implement \toolname in Isabelle/ML,
which allows it to be tightly integrated with
Isabelle's interactive environment.
\toolname first uses functions
from the Isabelle/ML ecosystem to
tokenize and parse its input (an entire apply-style proof)
into a custom AST.
Then,
\toolname crawls this AST to
replay each proof command
and collect the intermediate goals into a
second AST that represents Isar elements.
Finally,
\toolname processes this second AST into a string
that gets printed
to the output window as a clickable element.
When clicked,
the translated proof is automatically inserted into the proof document.\footnote{This is the same functionality used by Sledgehammer,
which allows the user to insert a proof found by an external solver \cite{hammering-away}.}
\toolname takes the input proof
in a pair of cartouches,
which lets the user specify exactly which portion of the proof should be translated:
\begin{lstlisting}[language=isabelle3]
lemma "[...]"
  using [...] unfolding [...] (* The user can decide whether to include these *)
apply2isar‹
  apply m1
  [...]
  by auto›
\end{lstlisting}

While Isabelle/HOL
can optionally produce proof terms,
they slow down proof checking and are disabled by default \cite{isabelle-proof-terms}.
\toolname does not work with proof terms or interact directly with the Isabelle kernel.
Instead, it replays each proof command by invoking the corresponding Isabelle/ML function.
This level of abstraction is appropriate for \toolname
because we intend it to produce structured Isar that faithfully
reflects the intermediate goals in the original apply-style proof,
which is data that is not evident in the final proof term.

\section{Implementation Challenges}\label{sec:implementation}
Isabelle is intentionally very flexible in both its procedural and declarative styles,
but there are significant structural differences between apply-style and structured Isar proofs.
\toolname supports the most frequently used proof commands;
\cref{tab:command-list} provides a full list of supported commands and their definitions.
In this section,
we explain important design decisions behind \toolname
and highlight some of the main challenges we encountered. 

\begin{table}
\begin{tabular}{c|l}\toprule
    {\isa |apply|} & Applies a method to the proof state, yielding a new proof state.\\ 
    {\isa |by|} & Similar to apply; ends a (sub)proof if the method closes all goals with backtracking.\\
    {\isa |done|} & Ends a proof if there are no more goals.\\
    {\isa |.|} & Immediate proof; ends proof if facts in proof state match goals (with unification).\\
    {\isa |..|} & Equivalent to {\isa |by standard|}.\\
    {\isa |using|} & Introduces facts into the proof state.\\
    {\isa |unfolding|} & Unfolds definitions in all goals and all facts in the proof state.\\
    {\isa |subgoal|} & Begins a subproof of the first goal in the proof state.\\
    {\isa |prefer|} & Moves a goal to the first position.\\
    {\isa |defer|} & Moves a goal to the last position.\\
    {\isa |back|} & Yields another possible state from the last method application.\\
    {\isa |supply|} & Introduces existing facts with an optional name.\\
    {\isa |proof|} & Begins an Isar proof for all current goals.\\
    {\isa |sorry|} & Proof placeholder; proves all current goals by cheating.\\\bottomrule
\end{tabular}
\caption{Proof commands that we handle.}
\label{tab:command-list}
\end{table}

\subsection{Operations on Multiple Goals}\label{sec:multiple-goals}

Some methods such as \isa{auto}, \isa{safe}, or \isa{simp_all}
can simultaneously affect multiple goals in the proof state.
For instance, consider:
\begin{lstlisting}[language=isabelle]
[...]         (* goals [ A1, B1, C1, D1 ] *)
   apply auto (* modifies A1 and C1; proves B1; does not modify D1 *)
[...]         (* goals [ A2, C2, D1 ] *) 
\end{lstlisting}
A simple way to handle this is to always print all goals
in the current proof state at every step.
This will even print goals that remain unchanged by a command:
\begin{lstlisting}[language=isabelle]
have "A2" "C2" "D1" by [...]
have "A1" "B1" "C1" "D1" by (auto) fact+ 
\end{lstlisting}
In this translation snippet,
the goal \isa{D1}
is carried over from the first \isa{have} to the second,
getting discharged the second time by \isa{fact+}.
This approach essentially treats the apply-style proof
as a linear sequence of proof state manipulations.
While this generates valid Isar that mimics the
structure of the original apply-style proof very closely,
it is very verbose 
and does not clearly indicate the \tit{purpose} of each command.
For instance,
consider the following example
where ten goals are discharged one at a time
(one may encounter such a scenario,
e.g., when proving locale instantiations):
\begin{lstlisting}[language=isabelle]
                  (* goals: [ A1, A2, A3, ..., A9, A10 ] *)
         apply m1 (* goals: [ A2, A3, ..., A9, A10 ] *)
        apply m2  (* goals: [ A3, ..., A9, A10 ] *)
       [...]
 apply m9         (* goals: [ A10 ] *)
by m10            (* goals: [ ] *)
\end{lstlisting}
Following the linear translation approach just described,
\toolname will generate:
\begin{lstlisting}[language=isabelle]
have "A10" by m10
have "A9" and "A10" by (m9) fact+
[...]
have "A2" and "A3" and [...] and "A9" and "A10" by (m2) fact+
show "A1" and "A2" and "A3" and [...] and "A9" and "A10" by (m1) fact+
\end{lstlisting}
There is a lot of unnatural repetition here---intuitively,
we might expect
\isa{m1} to prove \isa{A1},
\isa{m2} to prove \isa{A2},
and so forth.

Thus,
we implement a second approach,
which we call \isa{smart_goals},
that removes some of this redundancy by
only printing the goals that change after each command.
For instance,
\toolname can instead translate the previous example to:
\begin{lstlisting}[language=isabelle]
have "A10" by m10
have "A9" by m9
[...]
have "A2" by m2
have "A1" by m1
show "A1" and "A2" and [...] and "A9" and "A10" by fact+ (* Final combined show *)
\end{lstlisting}
As the goals are proven piecemeal,
\toolname combines them in a final \isa{show} line at the end of the Isar proof.\footnote{
In this example,
every \isa{have} could be replaced by \isa{show},
which would remove the need for the final combined \isa{show}.
However, this does not work in general:
\isa{m1} could have simultaneously proven
\isa{A1} along with some other non-top-level goal \isa{B},
in which case \isa{show "A1" and "B" by m1} would cause an error.
}

\subsection{Operations that Complicate Proof Flow}\label{sec:proof-flow}
In addition to manipulating goals,
commands may also change the proof state
by focusing a single goal or by introducing additional facts.

The current proof state when \toolname is invoked may already contain facts
introduced by \isa{using}, \isa{from}, etc.
As the structured Isar proof must be able to refer to those facts explicitly,
\toolname collects them and prints them as a named assumption
in the translation.

In an apply-style proof,
the \isa{supply} command allows one or more existing lemmas to be introduced
and given a local name.
For example:
\begin{lstlisting}[language=isabelle]
lemma "True" "P ⟹ P ∨ Q"
   supply myThms = TrueI disjI1 (* myThms = [ True, ?P ⟹ ?P ∨ ?Q ] *)
   apply (rule myThms(1))
  by (rule myThms(2)) 
\end{lstlisting}
The corresponding structured command for naming existing facts is \isa{note},
so \toolname converts the previous proof to:
\begin{lstlisting}[language=isabelle]
lemma "True" "P ⟹ P ∨ Q"
proof(-)
  note myThms = TrueI disjI1
  have "P ⟹ P ∨ Q" by (rule myThms(2))
  have "True" by (rule myThms(1))
  show "True" "P ⟹ P ∨ Q" by fact+ (* Final combined show *)
qed 
\end{lstlisting}
There is something peculiar here:
although the order of the original proof was reversed
(so that \isa{rule myThms(2)} got applied first),
the \isa{note} command and the corresponding \isa{supply} command are \tit{both}
at the beginning of the proof.
This is because if the \isa{note} command came after the two \isa{rule} applications,
\isa{myThms} would have been referenced before it was defined.
Our solution to this,
as seen in the prior example,
is to always lift generated \isa{note} commands to the beginning of the proof.
Note that the order of multiple \isa{note} commands must be preserved,
since later \isa{supply} commands can reference previous ones.

Finally, the \isa{subgoal} command focuses the first goal,
saving the remaining goals in a proof context.
When the subproof is finished,
the remaining goals are re-introduced from the proof context.
Additionally,
just as \isa{supply} introduces a local name for existing lemmas,
\isa{subgoal} can give a name to the local fact it proves.
For example:
\begin{lstlisting}[language=isabelle]
(* goals: [ P, P ∨ Q ] *)
subgoal myFact
  (* goals: [ P ] *)
  apply [...]
done
(* goals: [ P ∨ Q ] *)
by (rule disjI1[OF myFact])
\end{lstlisting}
\toolname handles the \isa{subgoal} command by opening a structured subproof:
\begin{lstlisting}[language=isabelle]
have myFact: "P"
proof(-)
  [...]
qed
\end{lstlisting}
Note that an unnamed \isa{subgoal}
can still be referenced as a fact literal.
Just as with \isa{supply},
this presents a fact-ordering problem:
na\"{i}vely reversing the order of the commands in the original proof
results in facts getting referenced before they are proven.
\toolname handles this
by moving all subproofs to the beginning of the Isar.
Because subproofs are standalone proofs,
we are not in danger of producing a broken proof,
so long as we maintain the order of \isa{subgoal} and \isa{supply} commands
(since a subproof or \isa{supply} may reference a previous \isa{subgoal} or \isa{supply} fact).
Since subproofs can be nested (and \isa{supply} commands can appear inside subproofs),
we recursively move subproofs (and \isa{note} commands) in the translation to the top of
the structured Isar proof block that contains them.

This solution produces logically valid Isar,
but it is not fully faithful to the logical flow of the original proof.
For greater readability,
one may want the original proof's flow to be captured in the translation.
Thus,
we implement a \isa{dummy_subproofs} option:
when enabled,
every moved subproof will be replaced with a ``dummy''
which proves the same statement by a \isa{fact} invocation.
The resulting translation remains more faithful to the original proof,
at the cost of being repetitive.
For instance,
with \isa{dummy_subproofs} enabled,
the previous example will yield the following Isar:
\begin{lstlisting}[language=isabelle]
have myFact: "P"
proof(-)
  [...]
qed
have "P ∨ Q" by (rule disjI1[OF myFact])
have "P" by fact (* Subproof dummy *)
show "P" "P ∨ Q" by fact+  (* Final combined show *)
\end{lstlisting}

\subsection{Interactions Between Commands}
In an apply-style proof, the
\isa{using} command introduces new facts to the proof state
after the goal has already been stated.
These facts can then be used in the next proof step,
such as the next method invocation via the \isa{apply} or \isa{by} commands.
The \isa{unfolding} command unfolds specified definitions
in all goals \tit{and} all facts in the proof state
(including those introduced by \isa{using}).
For example:
\begin{lstlisting}[language=isabelle]
lemma unfolding_and_using_ex:
  fixes P defines "R ≡ P"
  assumes 1: "P" and 2: "R ⟶ Q" shows "Q" "R"
  using 1 2       (* goals: [ Q, R ] | using: [ P, R ⟶ Q ] *)
  unfolding R_def (* goals: [ Q, P ] | using: [ P, P ⟶ Q ] *)
   apply simp     (* goals: [ P ]    | using: [] *)
  by (rule 1)
\end{lstlisting}
Because \isa{unfolding} can simultaneously affect all goals \tit{and} all facts
in the proof state,
it requires some care to handle,
particularly in its interaction with \isa{subgoal}.
We considered two approaches:
one that works generally
but introduces repetitiveness,
and one that avoids repetitiveness
but cannot be used when \isa{unfolding}
immediately precedes \isa{subgoal}.

The first approach is to treat \isa{unfolding} similarly to \isa{apply}:
for each \isa{unfolding} command,
\toolname creates a \isa{have} statement to capture that command's effect on the goals:
\begin{lstlisting}[language=isabelle]
proof(-)
  have "P" by (rule 1)
  have "Q" using 1 2 unfolding R_def by simp (* Repetitive unfolding after using *)
  have "R" unfolding R_def by fact+ (* Captures effects of unfolding on goals *)
  show "Q" and "R" by fact+ (* Combined show *)
qed
\end{lstlisting}
There is some repetition here: in the original proof,
\isa{unfolding R_def} is used after \isa{using 1 2},
so it must appear a second time in the translated proof
to capture its effects on facts \isa{1} and \isa{2}.
In an attempt to avoid this repetition, we implement a second
(perhaps more natural) approach under the option \isa{smart_unfolds}.
This approach groups the effects of
a sequence of \isa{unfolding}/\isa{using}
commands with the next \isa{apply} command.\footnote{
Or the next \isa{apply}-like command (\isa{by}, immediate proof (\isa{.}),
and default proof (\isa{..})).
}
For instance,
\toolname can translate the above example as follows,
where the effects of \isa{unfolding R_def}
are combined with those of \isa{apply simp}:
\begin{lstlisting}[language=isabelle]
proof(-)
  have "P" by (rule 1)
  show "Q" and "R" using 1 2 unfolding R_def by (simp) fact+
qed
\end{lstlisting}

Unfortunately,
this second approach cannot be used when \isa{unfolding}
immediately precedes \isa{subgoal};
because goals focused by \isa{subgoal} can be used later in a proof,
these goals must be present verbatim in the corresponding Isar.
Additionally,
if a \isa{using} command precedes the \isa{unfolding} command
(i.e.,
we have a \iisa{using-unfolding-subgoal} sequence),
then the facts introduced by \isa{using}
will be used in the first method application \tit{inside} the \isa{subgoal} subproof,
and we must still ensure those facts are properly unfolded.
Thus,
in this situation,
\toolname reverts to the first approach
since it correctly duplicates the \isa{unfolding}
steps in the translation.

\subsection{Shadowing}\label{sec:shadowing}
The \isa{subgoal} command,
introduced in \cref{sec:proof-flow},
skolemizes all meta-quantified variables
in the goal it focuses.
The user can use the \isa{for} keyword to choose names for some or all of the skolemized variables;
any variables left unnamed are given an internal representation
which cannot be explicitly referenced.
In structured Isar,
this corresponds to the \isa{fix} keyword.

Importantly, the user is allowed to pick names that shadow existing variables.
Shadowing is sometimes natural,
for example in induction proofs.
However, it can cause difficulties for \toolname
because Isabelle will use the same name for both the shadowed and the newly-skolemized variable
in the output window,
differentiated only by metadata (e.g., syntax highlighting).
For example:
\begin{lstlisting}[language=isabelle]
(* goals: [ ⋀x. P x y ]*)
subgoal for y
  (* goals: [ P y y ] (!!! These two y's are different!) *)
  apply [...] done
done
\end{lstlisting}
If we print \isa{have "P y y"} in the translation,
Isabelle interprets both occurrences of \isa{y} as the skolem variable.
Thus,
\toolname can ensure that fresh names are always used:
\begin{lstlisting}[language=isabelle]
have "⋀x. P x y"
proof(-)
  fix ya (* Renamed local variable *)
  [...]
  show "P ya y" by [...]
qed
\end{lstlisting}

Unfortunately,
this does not work in all cases.
To see why, suppose \isa{x} is an existing free variable in the proof context when \isa{subgoal} is invoked
in the following snippet:
\begin{lstlisting}[language=isabelle]
subgoal for x 
  by (simp add: some_lemma[where y=x])
\end{lstlisting}
As the newly-fixed variable \isa{x} is explicitly referenced
in the facts given to the \isa{simp} method,
renaming the skolemized \isa{x} would also require renaming it in the \isa{simp} invocation.
Doing this would require parsing each method in detail,
which we do not implement in \toolname because we
use the existing Isabelle/ML functions for parsing methods directly to the data
expected by the Isabelle/ML function for \isa{apply}.

Because this variable renaming behavior may need to be enabled or disabled
depending on the situation,
we implement the \isa{subgoal_fix_fresh} option.
By default,
this option is disabled,
which means
\toolname's default behavior can handle ``responsible'' shadowing,
where there are no goals containing the shadowed variable.
Enabling \isa{subgoal_fix_fresh}
lets \toolname handle ``irresponsible'' shadowing,
at the cost of failing when any renamed variables are explicitly referenced.

\subsection{Schematic Variables in Goals}\label{sec:schematic-variables}
In an Isabelle term,
\tit{schematic variables}
are variables which can be instantiated by higher-order unification \cite{isabelle-isar-reference-manual}.\footnote{
Technically,
a term with schematic variables is called a \tit{pattern} \cite{isabelle-isar-reference-manual}.
}
In inferences,
terms with meta-quantified variables (e.g., \isa{⋀x. P x})
are automatically converted into their schematic counterparts
(e.g., \isa{P ?x}) \cite{isabelle-isar-implementation}.
In an apply-style proof,
a method application may yield a schematic goal,
where the schematic variables represent unknowns that will be instantiated later in the proof.
For instance,
consider the following proof,
where the fact \isa{h1} becomes a schematic fact:
\begin{lstlisting}[language=isabelle]
lemma
  assumes h1: "⋀x y. P x ⟹ Q y" and h2: "P a" shows "Q b"
  (* h1: P ?x ⟹ Q ?y *)
  apply (rule h1) (* goals: [ P ?x ] *)
  by (rule h2) 
\end{lstlisting}
When \isa{rule h1} is applied,
Isabelle
unifies the conclusion of \isa{h1},
namely \isa{Q ?y},
with the goal,
namely \isa{Q b},
by instantiating \isa{?y} to \isa{b}.
However,
unification has no need to instantiate \isa{?x} here,
so it is left schematic and the new goal is \isa{P ?x}.
This goal can be discharged by instantiating \isa{?x}
to \tit{any} \isa{x} such that \isa{P x}.
In this case, instantiating \isa{?x = a} does the trick,
and unification again does this automatically when \isa{rule h2} is applied.

Schematic variables in goals are essentially free variables that will be instantiated by some future method application. 
However, \isa{have} does not allow schematic goals.
Thus,
\toolname avoids translating any portions of an apply-style proof
that contain schematic goals.
When a schematic goal appears,
\toolname collects all proof commands until
there are no more schematic goals.
In the generated Isar,
these collected commands are then printed as an apply-style proof
under a single \isa{have} statement.
As this leaves some apply-style proofs in the translation,
we consider such proofs ``partially translated'' in our evaluation.
We discuss possible alternative solutions in \cref{sec:future-and-related-work}.

\subsection{Types}\label{sec:types}
Until now,
we have been presenting examples without any type annotations.
While type inference often means type annotations are unnecessary,
this is not always the case.
In the following example,
the first lemma is not provable because \isa{x}
is only inferred to have type \isa|'a :: {zero,ord}|,
but the second lemma with type annotations is provable:
\begin{lstlisting}[language=isabelle2]
lemma not_provable: "∀x. 0 ≤ x" oops
lemma provable: "∀x :: nat. 0 ≤ x" by simp
\end{lstlisting}
While type information in an apply-style proof is maintained across proof states,
it is lost if terms are printed without annotations
(and the types cannot be fully recovered by inference).
To address this,
we provide the \isa{print_types} option.
When set to \isa{none},
\toolname will print no types;
when set to \isa{all},
\toolname will print all type annotations;
when set to \isa{necessary},
\toolname will only print type annotations when they are required.
\toolname checks if type annotations are required for a goal by printing it without type annotations,
re-parsing the result,
then checking if the re-parsed goal equals the original goal (modulo $\alpha$-equivalence).

\subsection{Proof Exploration and Partial Proofs}
While writing an apply-style proof,
a user might encounter goals which they would rather close with structured Isar.
Thus,
\toolname supports commands for proof exploration,
partial apply-style proof,
and mixed procedural-declarative proofs.

The \isa{defer} and \isa{prefer} commands can help with proof exploration
by reordering the set of goals \cite{isabelle-isar-reference-manual}.
These steps must be replayed as we process the original proof,
but they do not correspond to any structured steps because structured facts are explicitly
asserted and referenced. Thus, the translated proof will simply state the facts in
the new order with the appropriate labels without any explicit command being required.

The \isa{back} command also helps with proof exploration.
In an apply-style proof,
\isa{apply m}
applies the method \isa{m}
to obtain a sequence of potential next states
(or fails if \isa{m} could not refine the goal) \cite{isabelle-isar-reference-manual}.
By default,
the first state in the sequence is used,
but the user can request other possibilities with the \isa{back} command,
which backtracks and returns the next state in the sequence.
This is useful for proof exploration,
but is very brittle
and therefore explicitly banned in AFP entries.
An example of \isa{back} from the Isabelle tutorial is the following \cite{isabelle-tutorial}:
\begin{lstlisting}[language=isabelle]
lemma "⟦x = f x; T (f x) (f x) x⟧ ==> T x x x"  
  apply (erule ssubst) (* goals: [ T (f x) (f x) x ==> T (f x) (f x) (f x) ] *)
  back                 (* goals: [ T (f x) (f x) x ==> T x (f x) (f x) ]     *)
  back                 (* goals: [ T (f x) (f x) x ==> T (f x) x (f x) ]     *)
  back                 (* goals: [ T (f x) (f x) x ==> T x x (f x) ]         *)
  back                 (* goals: [ T (f x) (f x) x ==> T (f x) (f x) x ]     *)
  apply assumption     (* goals: []                                           *)
  done
\end{lstlisting}
\toolname automatically removes \isa{back} commands in its translation of this proof: 
\begin{lstlisting}[language=isabelle]
lemma "⟦ x = f x; T (f x) (f x) x ⟧ ⟹ T x x x"
proof(-)
  have "T (f x) (f x) x ⟹ T (f x) (f x) x" by (assumption)
  show "x = f x ⟹ T (f x) (f x) x ⟹ T x x x" by (erule ssubst) fact+
qed
\end{lstlisting}
This works because the \isa{by} command
is a terminal method application---it will
apply the given method(s) and automatically backtrack until a completely solved state is found
(or fail or time out if a solved state cannot be found).
Thus,
by making intermediate goals explicit,
\toolname can remove all \isa{back} commands since
the necessary backtracking is automatically handled by each \isa{by} command in the translation.

\toolname also supports the \isa{sorry} command,
which immediately proves all current goals by cheating---that is,
it is a proof placeholder.
This allows the user to close difficult goals in an apply-style proof with \isa{sorry},
then convert the proof to structured Isar before finishing it.
Additionally,
if a method application fails during our proof reconstruction,
we treat it instead as a \isa{sorry}.
This allows broken apply-style proofs to be partially converted to structured Isar.

Additionally,
the user can switch directly from an apply-style proofs to structured Isar:
\begin{lstlisting}[language=isabelle]
lemma "P" "Q" "R" (* goals: [ P, Q, R ] *)
    apply m1 (* goals: [ Q, R ]    *) 
proof(-)
  [...]
  show "Q" "R" [...]
qed
\end{lstlisting}
This pattern is considered brittle and is flagged by the Isabelle linter \cite{isabelle-linter}.
Once a structured Isar proof is opened,
it must prove all current goals---it is ``terminal'' in this sense,
like \isa{by}.
If a structured proof is initiated inside a \isa{subgoal} subproof
with the \isa{proof} command,
the associated \isa{qed} closes the subproof like \isa{by} and \isa{done}.
Structured Isar syntax is far more complex than apply-style syntax,
so \toolname avoids parsing and replaying such parts of a proof;
instead,
it simply assumes that it succeeds and copies it verbatim.
While \toolname maintains some of the user's formatting,
comments are unfortunately lost during tokenization.

In \toolname's proof replay,
we run the \isa{sorry} command to replicate the effects of
switching to Isar---this is necessary to ensure
that the proof replay correctly closes subproofs
that are concluded in the original apply-style proof with structured Isar.
Running \isa{sorry} in our proof replay is not problematic,
since the final generated proof is independently checked by the Isabelle kernel.
However,
it does mean that \toolname will dutifully reproduce even broken structured Isar
within an apply-style proof.
We view this side-effect of our implementation
as a feature that supports proof exploration and
partial proof translation.

\section{Evaluation}\label{sec:evaluation}
We tested \toolname with \ttt{Isabelle2025-2}.
We constructed our benchmark set by selecting thousands of
apply-style proofs from entries in
the \tit{Archive of Formal Proofs} (AFP).
Because many AFP entries contain relatively few apply-style proofs
(in part due to the AFP's general preference for structured proofs),
building our benchmark set by randomly sampling AFP entries would not be very effective.
Instead, we selected the five
AFP entries with the most occurrences of ``\ttt{apply}''---this
is a simple heuristic to identify entries that contain many
apply-style proofs.\footnote{
We simply \ttt{grep}'d for ``\ttt{apply}''
in each entry,
which counts spurious occurrences of ``\ttt{apply}''
such as those in definitions and lemma names.
However,
since we only need a rough heuristic,
this is acceptable.
}
These five entries are:
\ttt{Group-Ring-Module} \cite{Group-Ring-Module-AFP},
\ttt{AutoCorres2} \cite{AutoCorres2-AFP},
\ttt{Flyspeck-Tame} \cite{Flyspeck-Tame-AFP},
\ttt{Valuation} \cite{Valuation-AFP},
and \ttt{BNF\_Operations} \cite{BNF_Operations-AFP},
each of which contains multiple theory files.
As \ttt{AutoCorres2} is an extremely large entry,
we focus on the ten theory files that load in the primary session
with the most occurrences of ``\ttt{apply}'';
for all other entries, we test every theory file.
For each theory file that we tested,
we ran \toolname on every proof in the file containing at least
one \isa{apply} command.
We test the entirety of every proof,
including any \isa{unfolding} or \isa{using} commands
that may precede the first \isa{apply} command.
To aid in batch testing,
we wrote a wrapper in Isabelle/ML that processes an entire theory file
and wraps every apply-style proof
in \lstinline[language=isabelle3]{apply2isar‹[...]›} blocks.
Our final benchmark set contained 4461 apply-style proofs.
Our test bench had an AMD Ryzen 7 2700X 8-core CPU
and \qty{32}{\gibi\byte} of RAM.
We do not keep track of precise timing information
since our primary concern is the number of apply-style proofs that \toolname successfully
converts to structured Isar.
However,
in our evaluation,
we limited \toolname to a 30-second timeout to
ensure that it is reasonably performant.

\begin{figure}
\pgfplotstableread{ % data 
Label             TotalTranslations Partial Fails     
\ttt{Group-Ring-Module} 99.04     0.67    0.29 
\ttt{AutoCorres2}       89.92     5.46    4.62 
\ttt{BNF\_Operations}   27.68     70.06   2.26
\ttt{Flyspeck\_Tame}    67.20     32.44   0.36
\ttt{Valuation}         95.83     0.52    3.65
}\testdata
\begin{tikzpicture}
    \begin{axis}[
        y dir=reverse,
        enlarge x limits=false,
        hide x axis,
        xticklabel=\empty,
        height=0.20\textheight,
        width=0.84\textwidth,
        xbar stacked,   % Stacked horizontal bars
        xmin=0,         % Start x axis at 0
        ytick=data,     % Use as many tick labels as y coordinates
        ytick style={draw=none},
        legend style={
            draw=none, legend columns=-1,
            at={(axis cs:50,5)},anchor=north,
            /tikz/every even column/.append style={column sep=0.5cm}
        },
        yticklabels from table={\testdata}{Label}  % Get the labels from the Label column of the \datatable
    ]
    \addplot [draw=black, fill=green!80] table [x=TotalTranslations, meta=Label,y expr=\coordindex] {\testdata};   % "First" column against the data index
    \addplot +[pattern=crosshatch, pattern color=orange, draw=black] table [x=Partial, meta=Label,y expr=\coordindex] {\testdata};
    \addplot +[pattern=north east lines, pattern color=red!60, draw=black] table [x=Fails, meta=Label,y expr=\coordindex] {\testdata};
    \legend{Total Translations,Partial Translations,Fails}
    \end{axis}
\end{tikzpicture} 
\caption{A visualization of our evaluation data.
The solid green (leftmost) segment of each bar indicates the portion of complete translations;
the crosshatched orange (middle) segment indicates the portion of partial translations;
and the single-hatched red (rightmost) segment indicates the portion of failures.}
\label{fig:eval-data-bar-graph}
\end{figure}
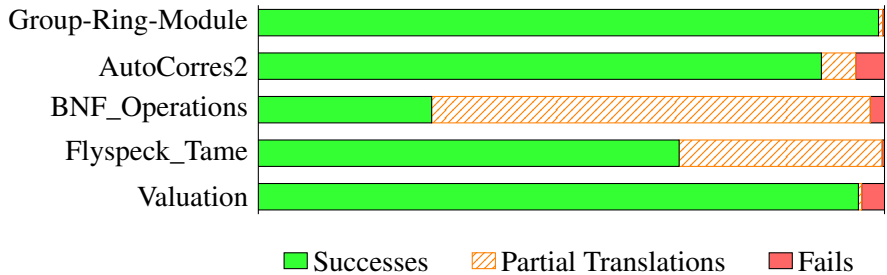

We present our evaluation results in \cref{tab:eval-data},
with a visualization in \cref{fig:eval-data-bar-graph}.
In \cref{tab:eval-data},
the ``\tbf{Entry}'' column indicates the AFP entry,
and the ``\tbf{Total Proofs}'' column indicates the total number of apply-style proofs
we tested in the corresponding entry.
We tested \toolname with our choice of default options:
\isa{named_facts},
\isa{smart_unfolds},
and \isa{smart_goals} enabled;
\isa{print_types} set to ``necessary'';
and all other options disabled.
The ``\tbf{Complete Translations}'' column indicates the number of proofs in the entry
that \toolname successfully translated to working Isar.

The number of proofs that were only partially translated to Isar due to schematic goals,
as discussed in \cref{sec:schematic-variables},
is indicated by the ``\tbf{Partial Translations}'' column.
We observed that most of \toolname's partial translations are still satisfactory.
Many partial translations involve extremely long apply-style proofs with
only short stretches of commands where schematic goals are present.
With these proofs,
the resulting translation is still primarily structured Isar,
with only certain steps requiring short apply-style proofs.
There were a few test cases where the majority of the proof contained schematic goals,
resulting in translations that change very little,
but these cases appear to be quite rare in our benchmark set.

\newcommand{\totalproofs}{\shortstack[l]{\tbf{Total}\\\tbf{Proofs}}}
\newcommand{\totaltranslations}{\shortstack[l]{\tbf{Complete}\\\tbf{Translations}}}
\newcommand{\partialtranslations}{\shortstack[l]{\tbf{Partial}\\\tbf{Translations}}}
\newcommand{\fails}{\multicolumn{3}{c}{\tbf{Fails}}}
\newcommand{\printfails}{\shortstack[l]{Print\\Fails}}
\newcommand{\timeoutfails}{\shortstack[l]{Timeout\\Fails}}
\newcommand{\miscfails}{\shortstack[l]{Misc.\\Fails}}
\newcommand{\standardopts}{\shortstack[l]{Standard\\Options}}
\newcommand{\otheropts}{\shortstack[l]{Other\\Options}}
\begin{table}
\begin{tabular}{lrrr|rrr}\toprule
                      &              &               &                      & \fails \\
    \tbf{Entry}       & \totalproofs & \totaltranslations    & \partialtranslations & \printfails & \timeoutfails & \miscfails\\\midrule
    \ttt{Group-Ring-Module} & 2387         & 2364          & 16                   & 2           & 3             & 2 \\
    \ttt{AutoCorres2}       & 952          & 856           & 52                   & 33          & 0             & 11 \\
    \ttt{Flyspeck-Tame}     & 561          & 377           & 182                  & 1           & 1             & 0 \\
    \ttt{Valuation}         & 384          & 368           & 2                    & 1           & 10            & 3 \\
    \ttt{BNF\_Operations}   & 177          & 49            & 124                  & 1           & 3             & 0 \\\bottomrule
\end{tabular}
\caption{Data from our evaluation of \toolname's complete translations,
partial translations, and failures.}
\label{tab:eval-data}
\end{table}

We observed that many of \toolname's failures are caused
by inconsistencies in Isabelle's term-printing mechanism.
In general,
printing goals and re-parsing them is not guaranteed to succeed in Isabelle;
the re-parsing may result in a different term or
simply cause a syntax error.
In fact,
this can occur even when manually copying goals from the output window,
so this is not a problem localized to our implementation.
While this is a rare occurrence,
we primarily observed it in our benchmark set with certain goals that use custom syntax.
We hope to see further Isabelle development toward improving this round-trip reliability,\footnote{
Cross-reference the following 2024 discussion on the Isabelle Users Mailing List:
\url{https://lists.cam.ac.uk/sympa/arc/cl-isabelle-users/2024-11/msg00006.html}.
}
since we do not see any general solution for this failure case
beyond addressing this underlying cause.
In \cref{tab:eval-data},
the ``\tbf{Fails}'' column indicates the number of proofs in the entry
for which \toolname did not produce working Isar---the
``Print Fails'' subcolumn indicates the number of failures
due to inconsistencies in Isabelle's term-printing mechanism;
the ``Timeout Fails'' subcolumn indicates the number of timeouts;
and the ``Misc. Fails'' subcolumn indicates the number of all other failures.
We observed that many of the timeouts are caused by extremely slow re-parsing of certain
goals
(as described in \cref{sec:types}, we re-parse goals
when \isa{print_types} is set to ``necessary'').
In these cases,
we observed that
manually copying the offending goals
into a \isa{have} statement resulted in extremely slow parsing by the \isa{have}
command itself,
so we do not see a general solution to these timeouts.
The sporadic miscellaneous failures do not appear to fit into a single category,
though we observed that they typically involve edge-case behavior of uncommon methods
that succeed in the original apply-style proof but fail in the translation.
Notably,
proof length did not seem to cause failures.
Indeed,
the longest apply-style proof successfully handled by \toolname
contained over 350 commands.

\section{Future and Related Work}\label{sec:future-and-related-work}
As discussed in \cref{sec:implementation},
translating apply-style proofs to structured Isar is far from straightforward.
Here,
we discuss some high-hanging fruit which could improve robustness
or provide additional convenience to the user.
We also examine related work and discuss other features that might
be useful to integrate into \toolname.

Our current handling of shadowed variables could be improved.
As discussed in \cref{sec:shadowing},
renaming variables in a \isa{subgoal} command to avoid shadowing
will break commands in the subproof that explicitly refer to the old name.
While the \isa{subgoal_fix_fresh} option
allows the user to disable this renaming,
it requires manual intervention and does not succeed in all cases.
Handling all potential edge cases would require
renaming user-chosen names as well as all explicit references to them;
but we expect implementing this robustly would require more intricate parsing of individual methods.
As it stands,
this problem only appears a handful of times in our benchmark set.

It would also be possible to improve our handling of schematic goals,
as discussed in \cref{sec:schematic-variables}.
One approach would be to obtain the instantiations made during the proof,
which could be done by examining proof terms.
Alternatively,
Isabelle provides the \isa{schematic_goal} theory command,
which is similar to \isa{lemma}
except that it allows goals to be schematic.
The final theorem produced by \isa{schematic_goal}
is thus determined by the instantiations made by the proof.
It is possible that this mechanism could be adapted to
allow \toolname to handle schematic goals more nicely.

Additionally,
we do not currently support custom commands.
While custom commands did not appear in our benchmark set,
some specialized libraries use them (e.g. \cite{Refine_Imperative_HOL-AFP,lammich-isabelle-sepref}).
While it would be difficult to support custom commands---which could do almost anything---in a principled way,
we could potentially support bypassing them (similar to our handling of schematic goals;
see \cref{sec:schematic-variables}).

In 2012,
Wiedijk developed a proof language called \ttt{miz3} for HOL Light
that combines the procedural and declarative styles
\cite{hol-light-procedural-declarative-synthesis}
based on the language for the Mizar system \cite{mizar-overview}.
Wiedijk also implemented \ttt{miz3\_of\_hol}, an automatic translation to \ttt{miz3} from standard
procedural HOL Light proofs.\footnote{The code for \ttt{miz3} and \ttt{miz3\_of\_hol} is included in the HOL Light distribution: \url{https://github.com/jrh13/hol-light/tree/master/miz3}.}
We intend for \toolname to convert procedural proofs to declarative ones
entirely within the existing Isabelle/Isar framework;
as structured Isar has been very widely adopted by Isabelle users,
we do not strive to develop a new declarative (or declarative-procedural hybrid) proof language.
While the high-level approach of \ttt{miz3\_of\_hol} is similar to our approach with \toolname,
HOL Light is designed to be a clean and simple system \cite{hol-light-overview},
and this makes the implementation of \ttt{miz3\_of\_hol} simpler than
that of \toolname;
as discussed in \cref{sec:implementation},
much of our implementation effort focused on
correctly handling the various Isabelle/Isar proof commands and their interactions.
For instance,
the \ttt{miz3\_of\_hol} implementation 
represents a procedural proof step using a datatype with two constructors
(to represent a possibly-nested tactic),
whereas our corresponding datatype uses fifteen constructors (to represent the various procedural proof commands in Isabelle/Isar).
Our proof replay must therefore handle these fifteen cases
and account for the various interactions between these cases.\footnote{
\cref{tab:command-list},
which lists the commands we handle,
only contains fourteen entries;
the additional constructor is to account for the possibility
of facts being introduced into the proof context (e.g., via \isa{using} or \isa{from})
prior to the invocation of \toolname,
as discussed in \cref{sec:proof-flow}.
}

In 2015,
Adams developed the Tactician tool for Rocq to pack (resp. unpack) chained sequences
of tactics into (resp. out of) a single proof step \cite{tactician}.
Similar ideas were explored by Magaud
to convert entire Rocq proof scripts to a single step \cite{toward-transformations-of-coq-scripts}.
In 2025,
Shi et al.\ explored ``deautomation'' in Rocq \cite{rocq-deautomation}---that is,
automatically refactoring a highly complex proof involving automated steps
into a simpler proof involving more ``primitive'' steps.
Notably,
they conducted a user study in which participants
expressed their desire for deautomation due to the
difficulties of working with heavily automated proofs.
As Isabelle also allows users to construct complex methods via combinators,
integrating deautomation into \toolname could prove useful.
However, deautomation relies on inspecting the behavior of individual methods.
Since Isabelle/HOL's automation is quite ``black-box'',
implementing such a feature could prove difficult.
Even splitting up a method sequence like \isa{apply (m1, m2)}
into two separate \isa{apply} commands is not necessarily straightforward
in light of possible interactions with other commands like \isa{using}.
However,
Kappelmann's recent Zippy tool,
which implements white-box automation
for Isabelle \cite{zippy-arxiv,Zippy-AFP},
may be helpful.

In 2022,
Megdiche et al.\ developed an Isabelle linter
which flags bad style
including (but not limited to)
complex steps (involving many combined methods),
complex initial and terminal Isar methods,
outdated methods and brittle commands (like \isa{back}),
and switching to structured Isar in an apply-style proof \cite{isabelle-linter}.
Note that the AFP automatically rejects submissions that do not pass certain linter checks.
The linter works syntactically on the proof document itself---unlike \toolname,
it does not involve replaying proofs or examining the proof state.
While it formerly had some automatic refactoring capability,
this feature only worked on syntactic anti-patterns and was ultimately removed.
One notable aspect of the Isabelle Linter
is its integration with the Isabelle/PIDE document model \cite{wenzel-isabelle-pide}.
Currently,
\toolname is implemented entirely within the Isabelle/ML ecosystem,
but better integration with the Isabelle/PIDE framework could improve the user experience.

Haftmann's Sketch and Explore tools
can print an Isar proof skeleton after a single method application,
optionally running a specified method or Sledgehammer on each of the new goals \cite{isabelle-sketch-and-explore}.
Building on this,
Tan et al.\ developed Super Sketch,
which aims to offer more fine-grained automation options \cite{super-sketch-and-super-fix}.
Both tools are intended for proof exploration,
and not for refactoring existing proofs.
In addition to Super Sketch,
Tan et al.\ also developed Super Fix,
a tool for making small proof fixes (e.g., one-line changes) in bulk.
Toward producing more natural translations,
it may be worthwhile to integrate a tool like Sketch into \toolname;
for instance,
specific methods in the apply-style proof,
such as \isa{induct},
could translate to a structured subproof whose skeleton is generated by Sketch.
However,
we believe this will require careful handling of multiple subgoals,
making the integration nontrivial.

As part of their AutoCorrode project,
AWS Labs has very recently launched Isabelle Assistant,
which integrates LLMs into the Isabelle interactive environment to perform many tasks
including proof refactoring \cite{AutoCorrode}.
As one of the Isabelle Assistant features,
the user can ask it to convert an apply-style proof to structured Isar.
Though we have not thoroughly evaluated Isabelle Assistant's effectiveness at this task,
our initial impression is that it can be slow and unreliable in certain situations.
For instance,
we found that Isabelle Assistant with Anthropic's Claude Opus $4.6$ LLM failed to convert the following proof
(taken from the Isabelle/HOL library)
to structured Isar:
\begin{lstlisting}[language=isabelle]
lemma (in monoid) division_weak_partial_order: (* From HOL-Algebra.Divisibility *)
  "weak_partial_order (division_rel G)"
  apply unfold_locales
        apply (simp_all add: associated_sym divides_antisym)
    apply (metis associated_trans)
   apply (metis divides_trans)
  by (meson associated_def divides_trans)
\end{lstlisting}
In this example,
the \isa{unfold_locales} invocation yields seven subgoals,
and the \isa{simp_all} invocation proves four of those subgoals and modifies the others.
The effects of the \isa{simp_all} invocation appear to have confused the LLM,
leading it to produce a broken proof.
On the other hand, \toolname can handle this proof without issue.
It is worth noting,
however,
that when Isabelle Assistant does succeed,
its translations are arguably more ``natural''
than those produced by \toolname.
Despite this,
we view the algorithmic approach of \toolname as a strong benefit---the output of \toolname is predictable and does not impose stylistic choices onto the user
(and as discussed in \cref{sec:motivating-example}, ``naturalness'' is a subjective metric).
\toolname also has the advantage that it can be run locally on moderate hardware,
rather than requiring a subscription to a cloud provider.
While we believe there are many virtues in taking an algorithmic approach, 
it may be interesting to explore
how LLM-based tools could
make the output of \toolname more natural and human-readable.
Notably,
the fact labels printed by \toolname
make the fact dependency tree of the structured proof explicit and follow a uniform naming scheme,
which we believe could be valuable information for an LLM-based tool.

\section{Conclusion}\label{sec:conclusion}
Our tool \toolname,
which translates apply-style proofs to structured Isar,
works well in practice and achieves our goal of
translating a wide range of apply-style proofs.
\toolname aims to improve the readability,
maintainability, and robustness of Isabelle/HOL developments.
We envision this being useful both in maintaining existing developments
and in supporting active proof developments.
We believe \toolname is a valuable addition to the Isabelle/HOL toolbox,
and we look forward to future improvements and new features.

\section{Acknowledgements}
This work was supported in part by the
Defense Advanced Research Projects Agency (DARPA) under contract FA8750-24-2-1001.
Any opinions, findings, and conclusions or recommendations
expressed here are those of the authors and do not necessarily reflect the views of DARPA.

Many thanks to Fabian Huch for his valuable feedback on the paper and
his Isabelle/ML guidance.
We also thank Andrew Marmaduke, Kartik Sabharwal, Shweta Rajiv, and Apoorv Ingle
for useful discussions and helpful feedback on the paper.
For help with specific technical questions via the Isabelle Zulip channel,
we thank Jonathan Juli\'{a}n Huerta y Munive, Kevin Kappelmann,
and Lukas Bartl.
Finally,
we thank the anonymous ITP reviewers for their valuable feedback and comments.

\bibliography{apply2isar}

\end{document}
\endinput